\documentclass{elsart}
\usepackage{natbib}
\usepackage{epsfig}
\usepackage{amssymb}

\begin{document}

\begin{frontmatter}

\title{The Complex Topology of Chemical Plants}

\author[1,2]{J. S. Andrade Jr.\corauthref{cor}},
\corauth[cor]{Corresponding author}
\ead{soares@fisica.ufc.br}
\author[2,3]{D. M. Bezerra},
\author[1,4]{J. Ribeiro Filho},
\author[1]{A. A. Moreira}

\address[1]{Departamento de F\'\i sica, Universidade Federal
do Cear\'a, 60451-970 Fortaleza, Cear\'a, Brazil}

\address[2]{Programa de
P\'os-Gradua\c c\~ ao em Engenharia Qu\'\i mica, Universidade Federal
do Cear\'a , 60451-970 Fortaleza, Cear\'a, Brazil}

\address[3]{Ger\^encia de
Otimiza\c c\~ ao, LUBNOR, PETROBRAS, Fortaleza, Cear\'a, Brazil}

\address[4]{Departamento de Matem\'atica, Universidade Estadual Vale do
Acara\'u, 62040-370 Sobral, Cear\'a, Brazil}

\begin{abstract}
We show that flowsheets of oil refineries can be associated to complex
network topologies that are scale-free, display small-world effect and
possess hierarchical organization. The emergence of these properties
from such man-made networks is explained as a consequence of the
currently used principles for process design, which include heuristics
as well as algorithmic techniques. We expect these results to be valid
for chemical plants of different types and capacities.
\end{abstract}


\end{frontmatter}

Several systems in nature and human society are constituted by a large
number of interacting agents that form a network with complex geometry
exhibiting small-world, scale-free and hierarchical features
\cite{Watts99,Barabasi02}. Examples include many of the existing 
metabolic processes \cite{Jeong00}, food webs \cite{Williams00}, 
and the relations in social groups \cite{Liljeros01}. An important
aspect always present in these systems is a self-regulatory behavior
that is responsible for their robustness to perturbations and
flexibility to respond to environmental changes and external
stimulus. Such a relation between robustness and flexibility, that
represents a necessary ingredient to ensure and maintain the
functionality of the system, generally results from the fact that the
global organization of agents is reached in the absence of any central
control, but as a collective phenomenon emerging from local
interactions. In this way, the generation of a decentralized structure
prevents the occurrence of those vulnerable points where an attack
could lead to the overall collapse of the network \cite{Cohen00}. 
At the same time, this non-central feature makes the complex system 
more adaptable, because it allows for the local response of agents 
to the variability of external conditions and eventual specific demands.

An obvious question arising from these facts and ideas that is far
from being answered is how local interactions among elementary units
of the network can naturally generate a dynamics with global
organization. At present, the study of these complex systems depends
on the development of new techniques to elucidate their behavior by
({\it i}) analysing the topological structure of their network of
interactions, and ({\it ii}) investigating the origin and
characteristics of their intrinsic self-organized dynamics. The
interplay between these two elements, namely, topology and dynamics,
can be very effective in revealing the mechanisms behind the
regulatory strategies of real systems in nature. Furthermore, this
could lead to novel ideas for design of artificial networks with
improved performance that display small-world and/or scale-free
properties like, for instance, the network of electrical energy
distribution \cite{Faloutsos99} and the Internet \cite{Vespignani04}.

In many ways, a chemical plant can be considered as a complex system
whose dynamics has to follow very strict conditions of robustness and,
on the other hand, to operate with certain flexibility. Specifically,
the project of a chemical process is the result of an enterprise
decision supported by a strategic plan to attend the market of a
region and generate profit. In the case of an oil refinery, for
example, this type of industry must be able to provide a broad
spectrum of chemicals that are produced at large amounts, but obeying
very strict specifications within a productive process that can also
be very peculiar in terms of safety and environmental impact. The
design of a chemical plant certainly represents one of the most
intricate and creative engineering activities \cite{Seider99}. A task
that is usually performed by chemical engineers, it requires a
detailed knowledge of the thermodynamics and transport phenomena
taking place at the level of each unitary operation that constitutes
the system. Furthermore, the designer must have a sufficiently
integrated view of the process to succeed in the development of its
complex flowsheet and selection of operational conditions, with also
enhanced performance in terms of selectivity and cost of the desired
product. 

Finally, it is of paramount importance to mention that the
accomplishment of a project for a chemical process is surely not a
one-to-one task, i.e., two effective solutions obtained from different
design teams might be substantially different. In order to overcome
the great complexity and expedite the development of the process
flowsheet, some {\it reasonable choices} have to be made that are not
formally based on mathematical or physical principles. Indeed, this
{\it non-determinism} represents a crucial step of the design of
chemical plants and involves the application of {\it heuristics} or
rules of thumb to undergo the steps in the so-called preliminary
process synthesis phase \cite{Seider99}. These rules, known to lead to
the near-optimal designs in relation to the selection of raw
materials, distribution of chemicals, and operations of separation,
heat-transfer and pressure-change, have been built from the large {\it
know-how} and experience accumulated during the design of many
projects in the past.
\begin{figure}
\begin{center}
\includegraphics[width=12cm]{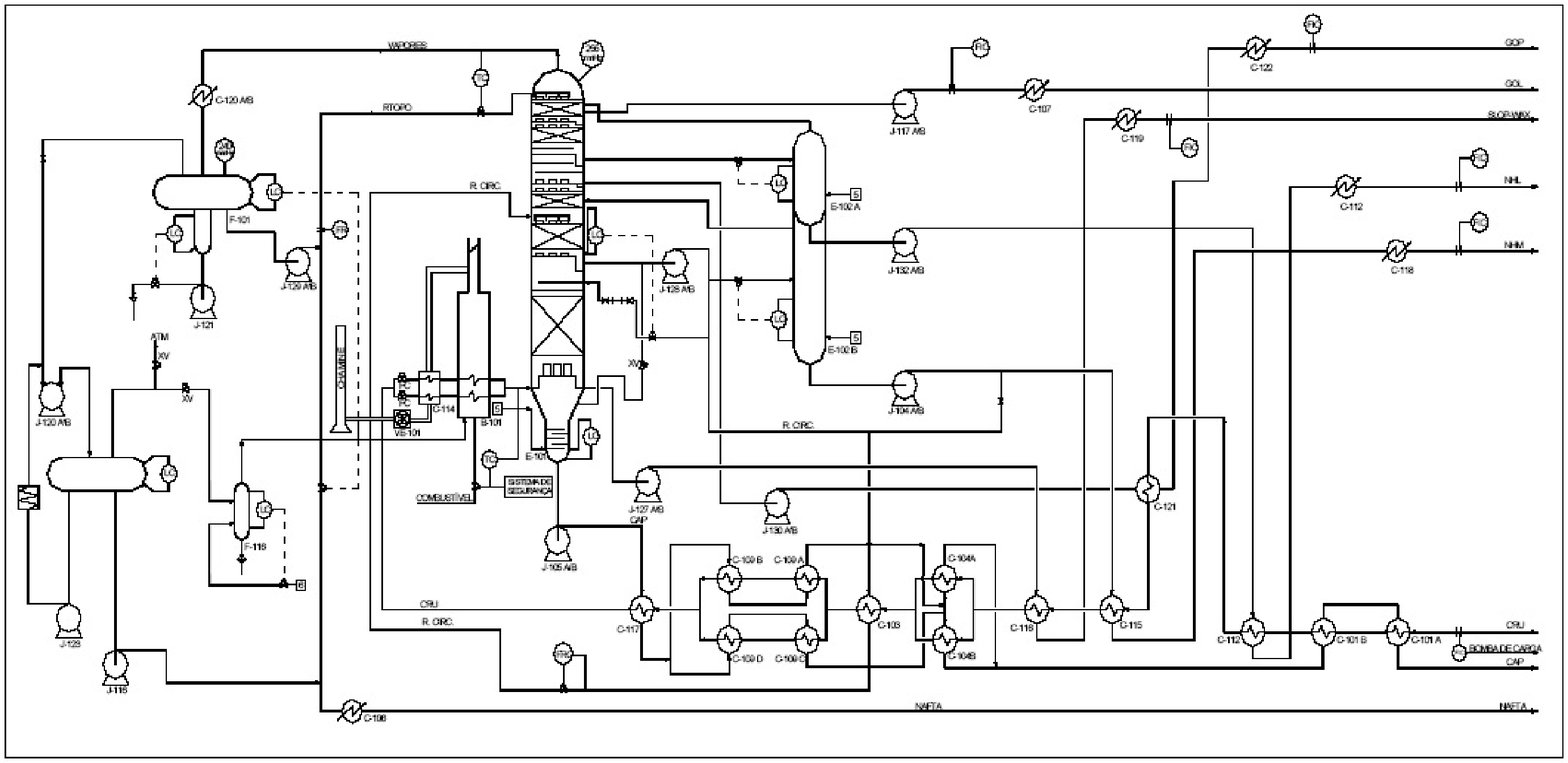}
\end{center}
\caption{Subset of a typical flowsheet showing the distillation unit 
of a chemical plant.}
\label{fig1}
\end{figure}
\begin{figure}
\begin{center}
\includegraphics[width=10cm]{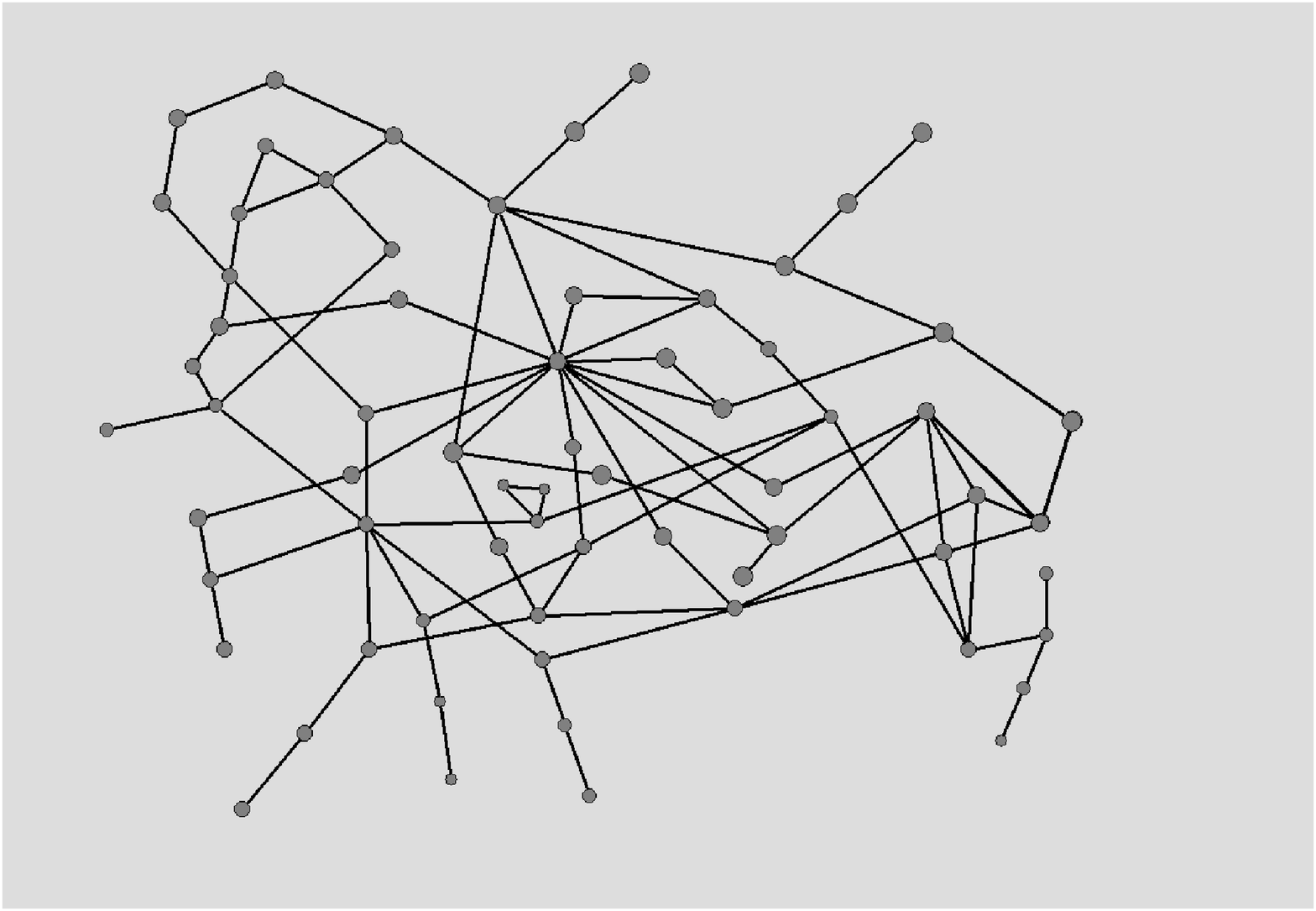}
\end{center}
\caption{Complex network representation of Fig.~1. The visualization 
is obtained by means of the Pajek code for large network analysis
\cite{Batagelj98}.}
\label{fig2}
\end{figure}

In the present study, we show evidence for the existence of a complex
topology behind the structure of chemical plants--typified here by oil
refineries--that is reminiscent of networks displaying scale-free and
hierarchical geometries. Our analysis is based on flowsheets of the
distillation and catalytic cracking units that constitute part of two
distinct oil refineries, owned and operated by PETROBRAS, the
Brazilian State--owned Oil Company. While the first chemical plant,
named here Refinery I, can process $8000~m^{3}/day$ of exclusively
light oil, the second one, or Refinery II, works with a mixture of
$80\%$ of heavy and $20\%$ of light oil and has a much larger
capacity, of approximately $30000~m^{3}/day$. In Fig.~1 we show the
subset of a typical flowsheet comprising some of the basic devices and
unitary process elements (e.g., valves, pumps, tanks, heat exchangers,
chemical reactors, and distillation columns, among others) connected
by a nested flow pipeline. The corresponding network shown in Fig.~2
is simply obtained by associating each of the process units displayed
in Fig.~1 to nodes, and their corresponding flow connections to
bonds. In Fig.~3, the complex structure of Refinery I is visualized
using the Pajek software for large network analysis \cite{Batagelj98}.
\begin{figure}
\begin{center}
\includegraphics[width=10cm]{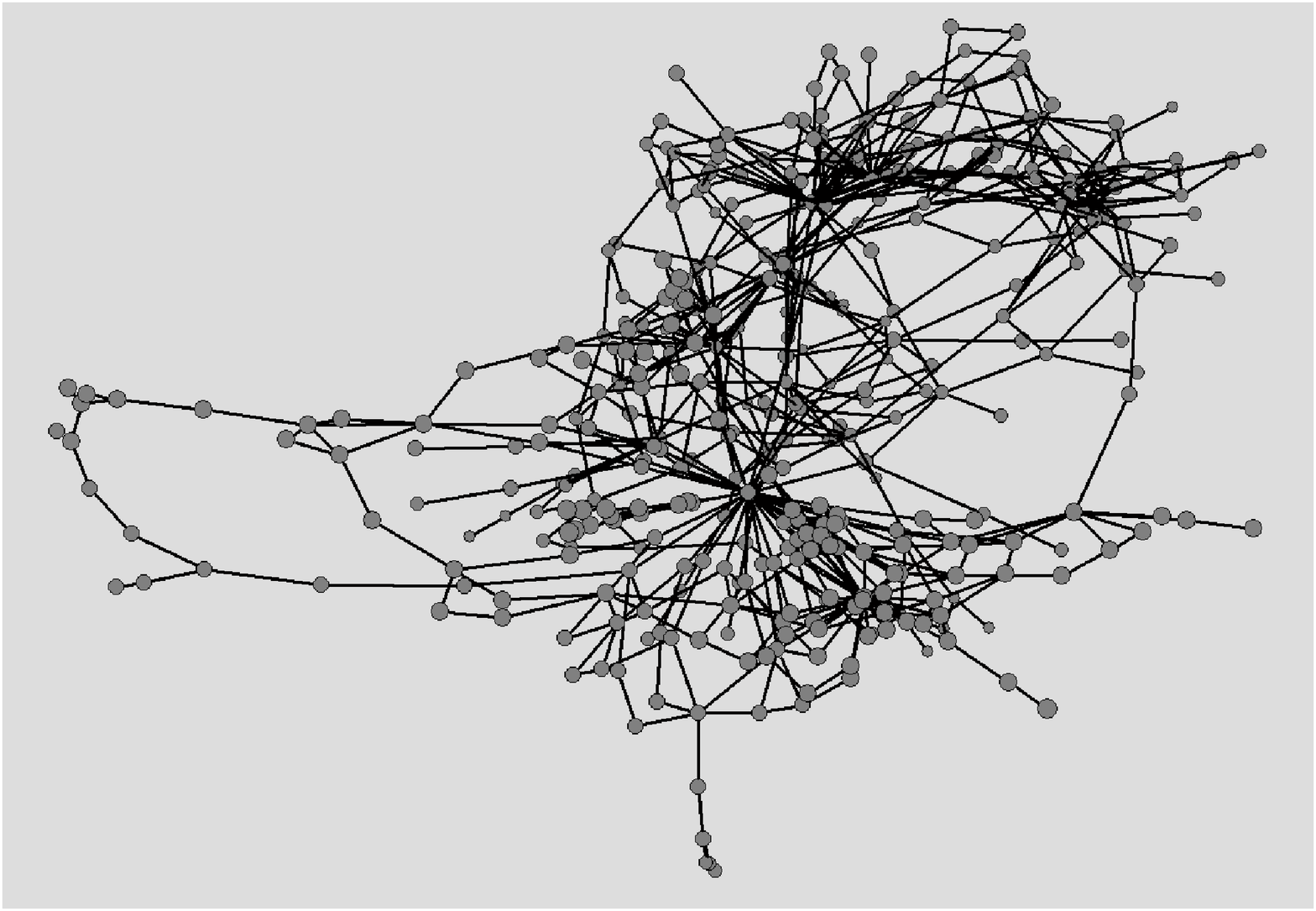}
\end{center}
\caption{Pajek visualization \cite{Batagelj98} of the complex network 
topology corresponding to the flowsheets of the distillation and
catalytic cracking units of Refinery I.}
\label{fig3}
\end{figure}

The results presented in Fig.~4 provide clear indication that the
network topology of Refinery I is scale-free \cite{Barabasi02}. More
precisely, its degree distribution for intermediate and large values
of the connectivity (degree) $k$ can be properly described as a
power-law, $p(k)\sim k^{-\gamma}$, with an exponent $\gamma=3.3 \pm
0.1$, as calculated from the least-square fit to the data in the
scaling region. The degree distribution for Refinery II follows an
entirely similar behavior, with an exponent $\gamma=3.2 \pm 0.2$. The
close agreement between these exponents can be substantiated by the
fact that the same set of design principles have been applied to the
project of both refineries, although their flowsheets are
substantially different and purpose-built to perform rather different
tasks.
\begin{figure}
\begin{center}
\includegraphics[width=10cm]{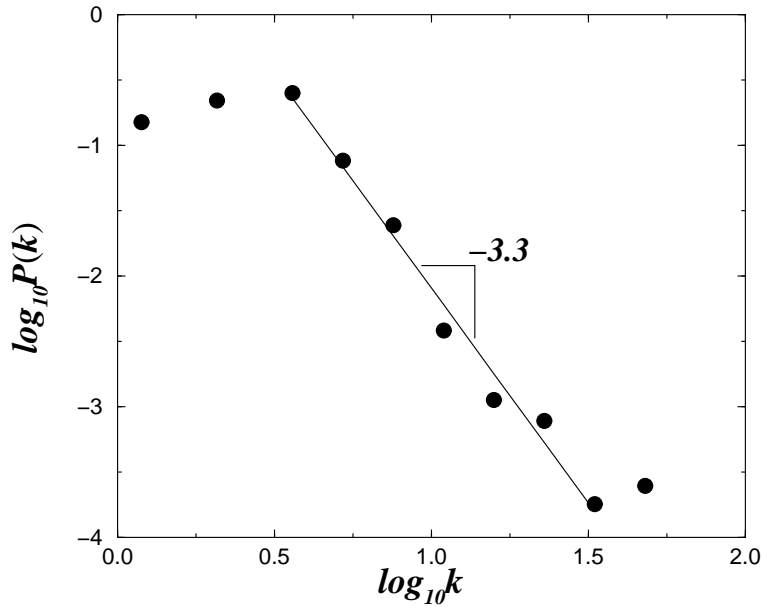}
\end{center}
\caption{Log-log plot showing the degree distribution of Refinery I.
The solid line corresponds to the best fit to the data in the scaling
region of a power-law, $p(k) \sim k^{-\gamma}$, with the exponent
$\gamma=3.3 \pm 0.1$.}
\label{fig4}
\end{figure}

Our calculation for the average length of the shortest path $\ell$
between two nodes, as defined in Ref.~\cite{Watts98}, gives $\ell=5.3$
and $5.9$ for Refinery I and II, respectively. When compared with the
corresponding sizes $N$ of the refineries, these small values of
$\ell$ suggest, but do not confirm, the presence of a {\it small-world
effect}. Strictly speaking, for a network to be considered {\it
small-world} \cite{Watts99,Amaral00,Doro04}, the shortest path $\ell$
should not only be small, but also grow slower than any positive power
of the system size $N$. Small-world networks also have large
clustering coefficient \cite{Watts98}. The clustering coefficient for
a node {\it i}, $C_{i}$, is defined as the fraction between the number
of connected pairs among its $k_{i}$ neighbors, $n_{i}$, and the
number of all possible connections between them,
\begin{equation}
C_{i} \equiv \frac{2n_{i}}{k_{i}(k_{i}-1)}~.
\label{clustering}
\end{equation}
Using the definition Eq.~(\ref{clustering}), we can compute an average
cluster coefficient, $C=\langle C_{i} \rangle$, for the entire
network. We find it to be $C=0.21$ and $0.16$ for Refinery I and II,
respectively. As shown in Table~1, these values are about $20$ times
larger than the clustering coefficients computed for random graphs,
$C_{\rm ran}=\langle k_{i} \rangle/N$, with the same average connectivity
$\langle k_{i} \rangle$, thus reinforcing our hypothesis that chemical
plants are small-world networks.
\begin{table}
\begin{center}
\includegraphics[width=10cm]{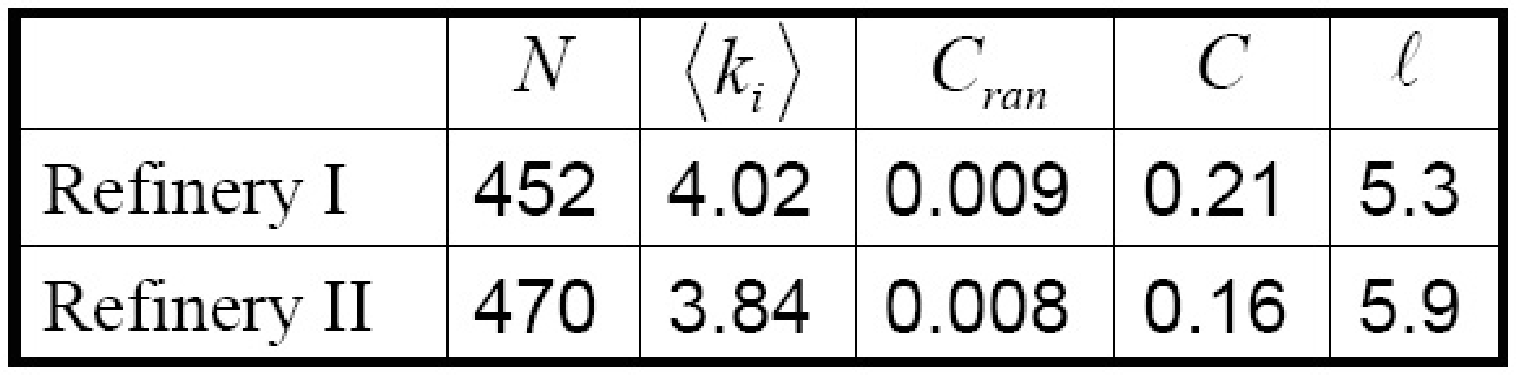}
\end{center}
\caption{Sizes and numerical values calculated for complex network 
parameters of Refineries I and II.} 
\end{table}
In order to show the hierarchical aspect of the refinery flowsheets, 
a more subtle type of analysis is required. Recently, it has been
analytically demonstrated that the intrinsic hierarchy of
deterministic scale-free networks leads to the scaling relation, $C(k)
\sim k^{-1}$, where $C(k)$ represents the average clustering
coefficient of nodes with connectivity $k$. Subsequently, the
occurrence of such a behavior has also been used as a way to identify
the hierarchical architecture of several real networks
\cite{Ravasz03}. Initially, the observation of $C(k)$ versus $k$
including {\it all nodes} in the data sets of the two refineries
revealed fluctuations that were large enough to impair any conclusive
description about the hierarchy of both network organizations. However, 
a detailed analysis of these numerical results indicates the presence 
of many nodes with $C_{i}=0$. These nodes mainly correspond to those
constituents of the flowsheets that have very low connectivity $k$
(e.g., valves, pumps, etc.) or are usually added in a secondary stage
of the design process to simply provide a single or multiple
connection among clusters of more specific unitary processes in the
system. As depicted in Fig.~5, the elimination of nodes with $C_{i}=0$
from the calculation of $C(k)$ for Refinery I enables us to identify a
scaling law, $C(k) \sim k^{-\beta}$ with $\beta=1.1 \pm 0.1$, that is
surprisingly close to the behavior expected for a hierarchical
topology. Adopting the same strategy for Refinery II we find an
entirely similar result, confirming the validity of our approach.
\begin{figure}
\begin{center}
\includegraphics[width=10cm]{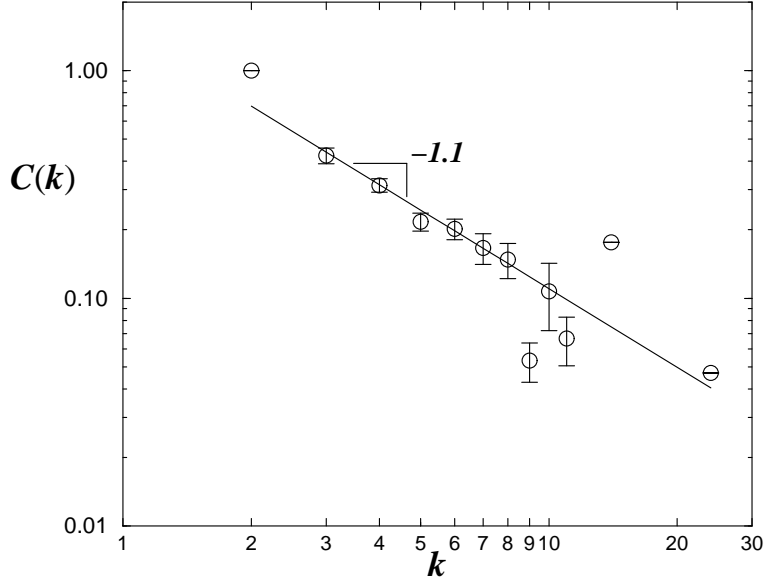}
\end{center}
\caption{Logarithmic plot of the clustering coefficient $C(k)$ against 
the connectivity $k$ for the network topology of Refinery I. The
straight line is the least-square fit to the data of a power-law, 
$C(k) \sim k^{-\beta}$, with the exponent $\beta=1.1 \pm 0.1$.}
\label{fig5}
\end{figure}

The identification of several networks with power-law degree
distribution in Nature has been generally justified in terms of their
evolution through a self-organized process in which hub elements are
spontaneously generated and represent the dominant parts of the
connected system \cite{Doro04}. Here we have shown that man-designed
networks composed by unitary processes and devices of a chemical
plant can also display scale-free behavior. Moreover, we found
significant evidence from real data to suggest that these networks
exhibit small-world effect and also have hierarchical organization in
their structure \cite{Ravasz03}. Although artificial, we believe that
these features stem from the process synthesis schemes tacitly adopted
in the design of chemical plants, which involves a combination of {\it
heuristics} and algorithmic techniques \cite{Seider99}. We expect these
results to be useful in the design stage as well as in the evaluation
and characterization of final flowsheets of refinery and other chemical
plants. 

We thank Diana Azevedo, C\'elio Cavalcante and Jo\~ao Augusto Paiva for
useful discussions, and CNPq, CAPES, and FUNCAP for financial support.

\end{document}